\documentstyle[11pt,IAU207_pasp,twoside,psfig]{article}
\def\edcomment#1{\iffalse\marginpar{\raggedright\sl#1\/}\else\relax\fi}
\reversemarginpar

\begin{document}
\title{A Comparative Study of Globular Cluster Systems in UGC 9799 and NGC 1129}
 \author{Myung Gyoon Lee, Eunhyeuk Kim}
\affil{Astronomy Program, SEES, Seoul National University, Seoul, 151-742, Korea}
\author{Doug Geisler}
\affil{Departamento de F{\'\i}sica, Grupo de Astronom{\'\i}a, Universidad
   de Concepci\'on, Casilla 160-C, Concepci\'on, Chile}
\author{Terry Bridges}
\affil{Anglo-Australian Observatory, P.O.Box 296, Epping, NSW 1710, Australia}
\author{Keith Ashman}
\affil{Department of Physic and Astronomy, University of Kansas, Lawrence, 
KS 66045-2151, USA}

\begin{abstract}
We present a preliminary analysis of 
HST-WFPC2 observations of globular cluster systems in the two
brightest galaxies, UGC 9799 (cD) and NGC 1129 (non-cD), located in the center of
rich clusters. 
\end{abstract}

UGC 9799 is a cD galaxy located in the center of the massive Abell 2052 cluster
at z=0.035, and is known to have the largest number of the globular clusters 
(N(total) $\approx 46,000$) 
and the highest specific frequency of globular clusters 
from the ground-based observations ($S_N = 20\pm6$, Harris, Pritchet, \& McClure 1995).
On the other hand, NGC 1129 is a giant, but not cD galaxy, located in the
center of a rich cluster AWM7 at z=0.018. Its globular cluster system has
not yet been studied.
The foreground reddenings are known to be $E(V-I)=0.051$ for UGC 9799 and
$E(V-I)=0.159$ for NGC 1129.
We adopt the redshift distance modulus $(m-M)_0=36.0$ for UGC 9799 and
$(m-M)_0=34.5$ for NGC 1129 based on the Hubble constant of $H_0 = 65 $ km/s/Mpc.

Deep images of these galaxies were obtained using the HST-WFPC2 
with $F555W$ ($V$) and $F814W$ ($I$) filters. 
We have obtained the photometry of the point sources in the images 
where bright galaxies were subtracted,
using the HSTphot  package (Dolphin 2000) and the image classification parameters.
Our photometry reaches $V \approx 27.2$ mag and $I \approx 26.0$ mag with 50 \% completeness.
Figure 1 displays the color-magnitude diagrams of the point sources in UGC 9799 and NGC 1129.
In Figure 1 there is seen a vertical structure at $0.8<(V-I)<1.5$ extending up to
$I \approx 23$ mag, which represents the globular clusters in these galaxies.
Faint blue objects are mostly background compact galaxies.

\begin{figure}
\centerline{
\psfig{figure=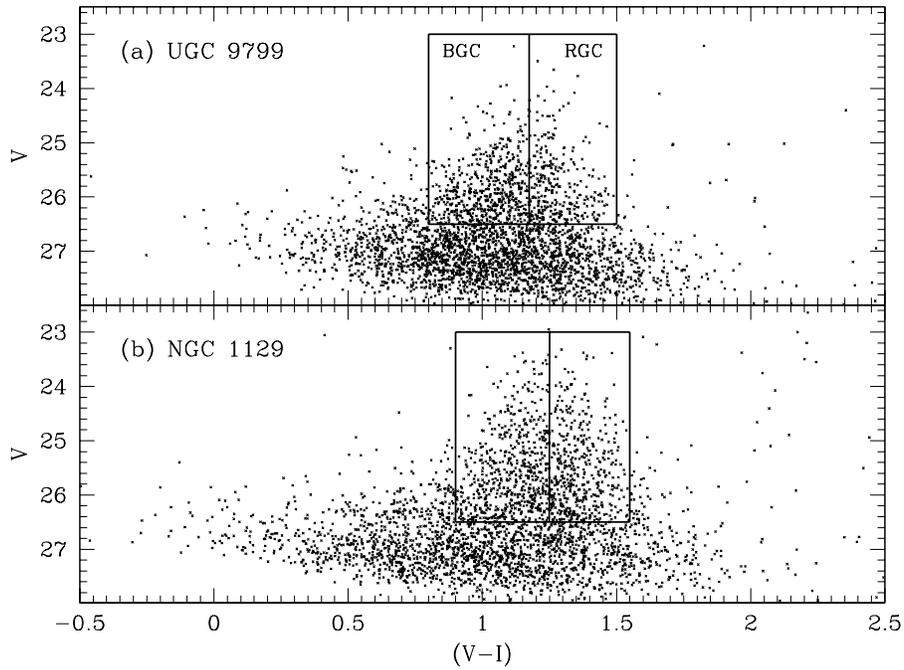,width=13cm,height=13cm} }
\vspace{-40truemm}
\caption{Color-magnitude diagrams of the point sources in
		UGC 9799 (a) and NGC 1129 (b). 
 The boxes represent the color and magnitude ranges 
for the bright blue and red globular cluster candidates with $V <26.5$ mag.}
\end{figure}

\begin{figure}
\centerline{
\psfig{figure=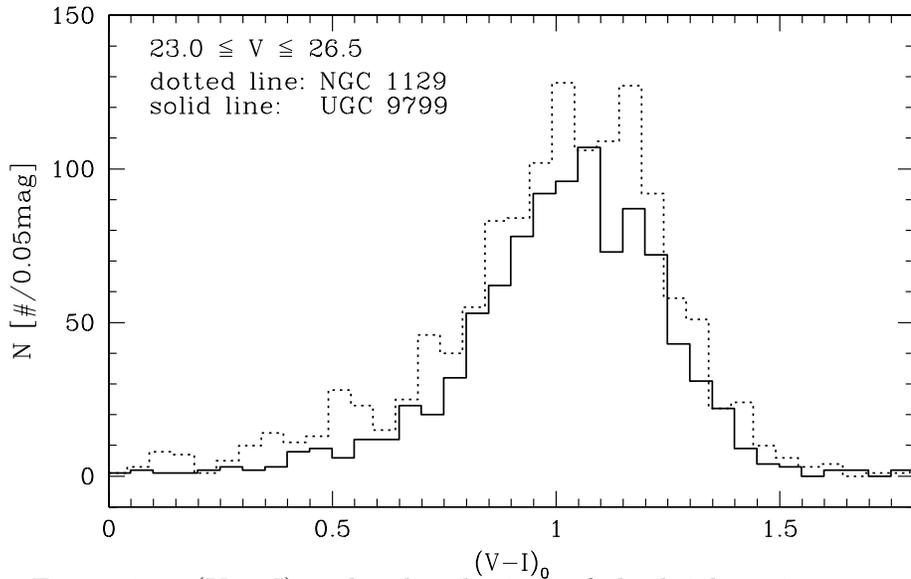,width=13cm,height=13cm} }
\vspace{-55truemm}
\caption{$(V-I)_0$ color distributions of 
the bright point sources (mostly globular clusters) 
with $V <26.5$ mag in UGC 9799 and NGC 1129}
\end{figure}

Figure 2 shows that the $(V-I)_0$ color distributions of the bright point sources 
with $V<26.5$ mag. In Figure 2 the dominant peaks are due to 
the globular clusters 
and the color distribution of the globular clusters in both galaxies 
are similarly bimodal: 
a blue peak at $(V-I)_0=1.07$ ([Fe/H] = --0.8) and a red peak at $(V-I)_0=1.17$ ([Fe/H] = --0.4) for UGC 9799, and
a blue peak at $(V-I)_0=1.02$ ([Fe/H] = --1.1) and a red peak at $(V-I)_0=1.17$ ([Fe/H] = --0.4) for NGC 1129.
The number of the bright globular clusters with $V<26.5$ mag we find is
860 for UGC 9799 and 1,060 for NGC 1129.
The ratio of the number of the blue globular clusters (BGC) and that of the red globular clusters (RGC) for UGC 9799 is derived to be 
N(BRC)/N(RGC) = 1.8, higher than that of NGC 1129,  N(BRC)/N(RGC) =1.3. 
 
The surface number density profiles of the globular clusters show that the globular clusters in both galaxies are spatially
more extended than those of the stellar halo, 
and the mean colors of the globular clusters are bluer than those of the stellar halo.
The RGCs are found to be more centrally concentrated than the RGCs in both galaxies.

Luminosity functions of the globular clusters (GCLFs) are derived after background
subtraction and incompleteness correction,
but they do not reach the turnovers which are expected to be at
$V\approx 28.7$ mag for UGC 9799 and $V\approx 27.5$ mag for NGC 1129.
We estimate the total number of the globular clusters from the GCLFs,
obtaining N(total)=$10,000\pm 700$ for UGC 9799 and N(total)=$7,000\pm 700$ for NGC 1129.
The total number of the globular clusters in UGC 9799 derived in this study is much
smaller than that derived from the ground-based observation by Harris et al. (1995).

From the integrated photometry of the galaxies the total magnitudes of the galaxies are estimated to be $V=12.10$ mag and $I=10.77$ mag for UGC 9799 ($r<63''$), 
and  $V=10.80$ mag and $I=9.41$ mag for NGC 1129 ($r<80''$).
Absolute total magnitudes of the galaxies are derived to be
$M_V=-24.02$ mag and $M_I=-25.30$ mag for UGC 9799, 
and $M_V=-24.08$ mag and $M_I=-25.32$ mag for NGC 1129, 
showing that both galaxies 
 belong to the brightest galaxies.

Finally we estimate the specific frequency of the globular clusters,
$S_N=N_t \times 10^{0.4(M_V +15)} = 2.5\pm 0.2$ for UGC 9799 and $S_N=1.7\pm 0.2$ for NGC 1129. These values are significantly lower than those for normal elliptical galaxies, 
and the value for UGC 9799 is much lower than that based 
on the ground-based observation (Harris et al. 1995).
If we use the total magnitudes of the galaxies given in the literature
($M_V=-23.4$ mag for UGC 9799, $M_V=-22.88$ mag for NGC 1129), we get 
$S_N=4.4\pm 0.3$ for UGC 9799 and $S_N=5.0\pm 0.5$ for NGC 1129.
This result is not consistent with 
the intracluster globular cluster model that
suggests that the globular clusters are not bound to individual galaxies, 
but bound to the gravitational potential of the clusters  (West et al. 1995).

\acknowledgments
This research is supported in part by the MOST/KISTEP International Collaboration Research Program (1-99-009).

\end{document}